\documentclass[11pt]{article}
\usepackage{amssymb,amsmath,amsfonts}
\usepackage{graphicx}
\usepackage{graphics}
\usepackage{eepic,epsfig}

\begin{document}

\title{Casimir densities for a plate in de Sitter spacetime }
\author{A. A. Saharian\thanks{%
E-mail: saharian@ictp.it }\, and T. A. Vardanyan \\
%EndAName
\textit{{Department of Physics, Yerevan State University} }\\
\textit{1 Alex Manoogian Street, 0025 Yerevan, Armenia }}
\maketitle

\begin{abstract}
Wightman function, the vacuum expectation values of the field squared and
the energy-momentum tensor are investigated for a scalar field with general
curvature coupling parameter in the geometry of a plate in the de Sitter
spacetime. Robin boundary condition for the field operator is assumed on the
plate. The vacuum expectation values are presented as the sum of two terms.
The first one corresponds to the geometry of de Sitter bulk without
boundaries and the second one is induced by the presence of the plate. We
show that for non-conformal fields the vacuum energy-momentum tensor is
non-diagonal with the off-diagonal component corresponding to the energy
flux along the direction perpendicular to the plate. In dependence of the
parameters, this flux can be either positive or negative. The asymptotic
behavior of the field squared, vacuum energy density and stresses near the
plate and at large distances is investigated.
\end{abstract}

\bigskip

\section{Introduction}

De Sitter (dS) spacetime is one of the simplest and most interesting
spacetimes allowed by general relativity. Quantum field theory in this
background has been extensively studied during the past two decades. Much of
early interest was motivated by the questions related to the quantization of
fields propagating on curved backgrounds. dS spacetime has a high degree of
symmetry and numerous physical problems are exactly solvable on this
background. The importance of this theoretical work increased by the
appearance of \ the inflationary cosmology scenario \cite{Lind90}. In most
inflationary models, an approximately dS spacetime is employed to solve a
number of problems in standard cosmology. During an inflationary epoch,
quantum fluctuations in the inflaton field introduce inhomogeneities which
play a central role in the generation of cosmic structures from inflation.
More recently, astronomical observations of high redshift supernovae, galaxy
clusters and cosmic microwave background \cite{Ries07} indicate that at the
present epoch the Universe is accelerating and can be well approximated by a
world with a positive cosmological constant. If the Universe would
accelerate indefinitely, the standard cosmology would lead to an asymptotic
dS universe. Hence, the investigation of physical effects in dS spacetime is
important for understanding both the early Universe and its future. Another
motivation for investigations of dS based quantum theories is related to the
holographic duality between quantum gravity on dS spacetime and a quantum
field theory living on boundary identified with the timelike infinity of dS
spacetime \cite{Stro01}.

The Casimir effect is among the most striking macroscopic manifestations of
non-trivial properties of the quantum vacuum. The presence of reflecting
boundaries alters the zero-point modes of a quantized field and shifts the
vacuum expectation values of quantities, such as the energy density and
stresses. As a result, vacuum forces arise acting on constraining
boundaries. The particular features of these forces depend on the nature of
the quantum field, the type of spacetime manifold, the boundary geometry and
the specific boundary conditions imposed on the field. Since the original
work by Casimir \cite{Casi48} many theoretical and experimental works have
been done on this problem (see, e.g., \cite{Most97}-\cite{Klim09} and
references therein).

The Casimir effect can be viewed as a polarization of vacuum by boundary
conditions. The interaction of fluctuating quantum fields with background
gravitational fields gives rise to another type of vacuum polarization (see,
for instance, \cite{Birr82,Grib94}). Here we will study an example with both
types of vacuum polarization. Namely, we evaluate the vacuum expectation
values for the field squared and the energy-momentum tensor of a scalar
field with general curvature coupling parameter induced by a plane boundary
on background of $(D+1)$-dimensional dS spacetime. Previously the Casimir
effect on background of dS spacetime described in planar coordinates was
investigated in Refs. \cite{Seta01} for a conformally coupled massless
scalar field. In this case the problem is conformally related to the
corresponding problem in Minkowski spacetime and the vacuum characteristics
are generated from those for the Minkowski counterpart multiplying by the
conformal factor. In particular, for the geometry of a single plate the
vacuum expectation value of the energy-momentum tensor vanishes. In \cite%
{Saha04} the vacuum expectation value of the energy--momentum tensor for a
conformally coupled scalar field is investigated in dS spacetime with static
coordinates in presence of curved branes on which the field obeys the Robin
boundary conditions with coordinate dependent coefficients (for
investigations of the Casimir energy in braneworld models with dS branes see
Refs. \cite{Noji00}). In these papers the conformal relation between dS and
Rindler spacetimes and the results for the Rindler counterpart \cite{Saha02}
were used. More recently, the Casimir densities in dS spacetime with
toroidally compactified spatial dimensions were investigated in \cite{Saha08}%
.

The present paper is organized as follows. In the next section the geometry
of the problem is described and the corresponding positive frequency
Wightman function is evaluated for the general case of the curvature
coupling parameter and for Robin boundary condition on the plane boundary.
In quantum field theory on curved backgrounds among the important quantities
describing the local properties of a quantum field and quantum back-reaction
effects are the expectation values of the field squared and the
energy-momentum tensor. In sections \ref{sec:Phi2} and \ref{sec:EMT}, by
using the expression for the Wightman function, we investigate the parts in
these expectation values induced by the plate. Simple asymptotic formulae
are obtained for small and large proper distances from the plate with
respect to the dS curvature scale. The special cases of Dirichlet and
Neumann boundary conditions are discussed in section \ref{sec:Dir}. The main
results are summarized in section \ref{sec:Conc}.

\section{Wightman function}

\label{sec:WF}

For a scalar field with curvature coupling parameter $\xi $\ the field
equation has the form%
\begin{equation}
\left( \nabla _{l}\nabla ^{l}+m^{2}+\xi R\right) \varphi =0,  \label{fieldeq}
\end{equation}%
where $\nabla _{l}$ is the covariant derivative operator and $R$ is the
Ricci scalar for the background spacetime. The values of the curvature
coupling parameter $\xi =0$ and $\xi =\xi _{D}\equiv (D-1)/4D$, with $D$
being the number of spatial dimensions, correspond to the most important
special cases of minimally and conformally coupled fields. In the present
paper the background geometry is the $(D+1)$-dimensional de Sitter
spacetime. We write the corresponding line element in planar coordinates
most appropriate for cosmological applications:%
\begin{equation}
ds^{2}=dt^{2}-e^{2t/\alpha }\sum_{i=1}^{D}(dz^{i})^{2}.  \label{ds2deSit}
\end{equation}%
The Ricci scalar and the corresponding cosmological constant are related to
the parameter $\alpha $ in the scale factor by the formulae%
\begin{equation}
R=D(D+1)/\alpha ^{2},\;\Lambda =D(D-1)/2\alpha ^{2}.  \label{Ricci}
\end{equation}%
In addition to the synchronous time coordinate $t$ it is convenient to
introduce the conformal time in accordance with $\tau =-\alpha e^{-t/\alpha
} $, $-\infty <\tau <0$. In terms of this coordinate the line element takes
the conformally flat form:%
\begin{equation}
ds^{2}=\alpha ^{2}\tau ^{-2}[d\tau ^{2}-\sum_{i=1}^{D}(dz^{i})^{2}].
\label{ds2Dd}
\end{equation}

We will assume the presence of a plate located at $z^{D}=0$ on which the
field obeys the Robin boundary condition%
\begin{equation}
\left( 1+\beta \partial _{z^{D}}\right) \varphi =0,\;z^{D}=0,  \label{BC}
\end{equation}%
with a constant coefficient $\beta $. Robin type conditions are an extension
of Dirichlet and Neumann boundary conditions and appear in a variety of
situations, including the considerations of vacuum effects for a confined
charged scalar field in external fields, spinor and gauge field theories,
quantum gravity and supergravity. Robin boundary conditions with
coefficients related to the curvature scale of the background spacetime
naturally arise for scalar and fermion bulk fields in braneworld models. In
this paper we are interested in the boundary induced effects on the vacuum
expectation values (VEVs) of the field squared and the energy-momentum
tensor assuming that the field is prepared in the Bunch-Davies vacuum state
\cite{Bunc78}. These VEVs are obtained from the corresponding Wightman
function in the coincidence limit of the arguments. In addition, the
Wightman function determines the response of the particle detector of
Unruh-DeWitt type, moving through the vacuum under consideration (see, for
instance, \cite{Birr82}).

In order to evaluate the Wightman function we employ the mode-sum formula
\begin{equation}
W(x,x^{\prime })=\langle 0|\varphi (x)\varphi (x^{\prime })|0\rangle
=\sum_{\sigma }\varphi _{\sigma }(x)\varphi _{\sigma }^{\ast }(x^{\prime }),
\label{Hadam1}
\end{equation}%
where $\left\{ \varphi _{\sigma }(x),\varphi _{\sigma }^{\ast }(x)\right\} $
is a complete set of solutions to the classical field equation satisfying
the boundary condition (\ref{BC}). The eigenfunctions realizing the
Bunch-Davies vacuum state and obeying the boundary condition have the form
\begin{equation}
\varphi _{\sigma }(x)=C_{\sigma }\eta ^{D/2}H_{\nu }^{(1)}(K\eta )\cos
[k_{D}z^{D}+\alpha (k_{D})]e^{i\mathbf{k}\cdot \mathbf{z}},\;\sigma =(%
\mathbf{k},k_{D}),  \label{eigfuncD}
\end{equation}%
with the notations $\eta =|\tau |$, $K=\sqrt{k^{2}+k_{D}^{2}}$,%
\begin{equation}
\mathbf{k}=(k_{1},\ldots ,k_{D-1}),\;\mathbf{z}=(z^{1},\ldots ,z^{D-1}).\;
\label{kD1D2}
\end{equation}%
The order of the Hankel function $H_{\nu }^{(1)}(x)$ in (\ref{eigfuncD}) is
given by the expression%
\begin{equation}
\nu =\left[ D^{2}/4-D(D+1)\xi -m^{2}\alpha ^{2}\right] ^{1/2},  \label{knD}
\end{equation}%
and the function $\alpha (k_{D})$ is defined by the relation
\begin{equation}
\;e^{2i\alpha (k_{D})}=\frac{i\beta k_{D}-1}{i\beta k_{D}+1}.
\label{alphakD}
\end{equation}%
With this definition the function (\ref{eigfuncD}) satisfies the boundary
condition at $z^{D}=0$. Note that we also have the formulae%
\begin{equation}
\cos [\alpha (k_{D})]=\frac{k_{D}\beta }{\sqrt{1+k_{D}^{2}\beta ^{2}}}%
,\;\sin [\alpha (k_{D})]=\frac{1}{\sqrt{1+k_{D}^{2}\beta ^{2}}}.
\label{alphakD2}
\end{equation}

The coefficient $C_{\sigma }$ in the expression for the eigenfunctions is
found from the orthonormalization condition
\begin{equation}
-i\int d\mathbf{z}\int_{0}^{\infty }dz^{D}\sqrt{|g|}g^{00}\left[ \varphi
_{\sigma }(x)\partial _{\tau }\varphi _{\sigma ^{\prime }}^{\ast
}(x)-\varphi _{\sigma ^{\prime }}^{\ast }(x)\partial _{\tau }\varphi
_{\sigma }(x)\right] =\delta (\mathbf{k-k}^{\prime })\delta
(k_{D}-k_{D}^{\prime }).  \label{normcond}
\end{equation}%
By using the Wronskian relation for the Hankel functions, it can be seen that%
\begin{equation}
H_{\nu }^{(1)}(K\eta )H_{\nu ^{\ast }}^{(2)\prime }(K\eta )-H_{\nu ^{\ast
}}^{(2)}(K\eta )H_{\nu }^{(1)\prime }(K\eta )=-\frac{4ie^{-(\nu -\nu ^{\ast
})\pi i/2}}{\pi K\eta }.  \label{Hankel}
\end{equation}%
In addition, by making use of the definition of the function $\alpha (k_{D})$%
, we can derive the integration formula%
\begin{equation}
\int_{0}^{\infty }dz^{D}\cos [k_{D}z^{D}+\alpha (k_{D})]\cos [k_{D}^{\prime
}z^{D}+\alpha (k_{D}^{\prime })]=\frac{\pi }{2}\delta (k_{D}-k_{D}^{\prime
}).  \label{NormInt}
\end{equation}%
Combining the last two relations with (\ref{normcond}), for the
normalization coefficient we find%
\begin{equation}
C_{\sigma }^{2}=\frac{e^{(\nu -\nu ^{\ast })\pi i/2}}{2(2\pi \alpha )^{D-1}}.
\label{normCD}
\end{equation}

Substituting the eigenfunctions (\ref{eigfuncD}) with the normalization
coefficient (\ref{normCD}) into the mode-sum formula (\ref{Hadam1}) for the
Wightman function, one finds
\begin{eqnarray}
W(x,x^{\prime }) &=&\frac{8(\eta \eta ^{\prime })^{D/2}}{(2\pi )^{D+1}\alpha
^{D-1}}\int dk\,e^{i\mathbf{k}\cdot (\mathbf{z}-\mathbf{z}^{\prime
})}\int_{0}^{\infty }du\,\cos [uz^{D}+\alpha (u)]\cos [uz^{D\prime }+\alpha
(u)]  \notag \\
&&\times K_{\nu }(\eta \sqrt{k^{2}+u^{2}}e^{-\pi i/2})K_{\nu }(\eta ^{\prime
}\sqrt{k^{2}+u^{2}}e^{\pi i/2}).  \label{WFSingle}
\end{eqnarray}%
where we have written the Hankel functions in terms of the modified Bessel
function $K_{\nu }(z)$. The formula (\ref{WFSingle}) can also be presented
in an equivalent form
\begin{eqnarray}
W(x,x^{\prime }) &=&W_{\mathrm{dS}}(x,x^{\prime })+\frac{4(\eta \eta
^{\prime })^{D/2}}{(2\pi )^{D+1}\alpha ^{D-1}}\int d\mathbf{k}\,e^{i\mathbf{k%
}\cdot (\mathbf{z}-\mathbf{z}^{\prime })}\int_{0}^{\infty }du\,\cos
[u(z^{D}+z^{D\prime })+2\alpha (u)]  \notag \\
&&\times K_{\nu }(\eta \sqrt{k^{2}+u^{2}}e^{-\pi i/2})K_{\nu }(\eta ^{\prime
}\sqrt{k^{2}+u^{2}}e^{\pi i/2}),  \label{WFSingle1}
\end{eqnarray}%
where%
\begin{eqnarray}
W_{\mathrm{dS}}(x,x^{\prime }) &=&\frac{4(\eta \eta ^{\prime })^{D/2}}{(2\pi
)^{D+1}\alpha ^{D-1}}\int d\mathbf{k}\,e^{i\mathbf{k}\cdot (\mathbf{z}-%
\mathbf{z}^{\prime })}\int_{0}^{\infty }du\,\cos [u(z^{D}-z^{D\prime })]
\notag \\
&&\times K_{\nu }(\eta \sqrt{k^{2}+u^{2}}e^{-\pi i/2})K_{\nu }(\eta ^{\prime
}\sqrt{k^{2}+u^{2}}e^{\pi i/2}),  \label{WF0}
\end{eqnarray}%
is the Wightman function for the dS spacetime without boundaries and the
second term on the right is induced by the plate at $z^{D}=0$. Two-point
function (\ref{WF0}) is well investigated in the literature \cite%
{Bunc78,Cand75,Dowk76,Bros96,Bous02} (see also \cite{Birr82}) and can be
presented in the form%
\begin{equation}
W_{\mathrm{dS}}(x,x^{\prime })=\frac{\alpha ^{1-D}\Gamma (D/2+\nu )\Gamma
(D/2-\nu )}{2^{(D+3)/2}\pi ^{(D+1)/2}\left( u^{2}-1\right) ^{(D-1)/4}}P_{\nu
-1/2}^{(1-D)/2}(u),  \label{WFdS}
\end{equation}%
where $P_{\nu }^{\mu }(u)$ is the associated Legendre function of the first
kind and
\begin{equation}
u=-1+\frac{\sum_{l=1}^{D}(z^{l}-z^{\prime l})^{2}-(\eta -\eta ^{\prime })^{2}%
}{2\eta \eta ^{\prime }}.  \label{u}
\end{equation}%
An alternative form is obtained by using the relation between the associated
Legendre function and the hypergeometric function (see, for example, \cite%
{Abra64}).

In order to transform the boundary induced part in (\ref{WFSingle1}) into a
more convenient form we write%
\begin{equation}
2\cos [u(z^{D}+z^{D\prime })+2\alpha (u)]=e^{iu(z^{D}+z^{D\prime })}\frac{%
i\beta u-1}{i\beta u+1}+e^{-iu(z^{D}+z^{D\prime })}\frac{i\beta u+1}{i\beta
u-1}.  \label{relcos}
\end{equation}%
Now, in the boundary induced part of (\ref{WFSingle1}) we rotate the
integration contour over $u$ by the angles $\pi /2$ and $-\pi /2$ for the
integrals with the first and second terms on the right-hand side of (\ref%
{relcos}), respectively. Assuming that $\beta \leqslant 0$, the Wightman
function is presented in the form%
\begin{eqnarray}
W(x,x^{\prime }) &=&W_{\mathrm{dS}}(x,x^{\prime })+\frac{(\eta \eta ^{\prime
})^{D/2}}{2(2\pi )^{D}\alpha ^{D-1}}\int d\mathbf{k}\,e^{i\mathbf{k}\cdot (%
\mathbf{z}-\mathbf{z}^{\prime })}\int_{k}^{\infty
}du\,e^{-u(z^{D}+z^{D\prime })}\frac{\beta u+1}{\beta u-1}  \notag \\
&&\times \left\{ K_{\nu }(\eta y)\left[ I_{-\nu }(\eta ^{\prime }y)+I_{\nu
}(\eta ^{\prime }y)\right] +\left[ I_{\nu }(\eta y)+I_{-\nu }(\eta y)\right]
K_{\nu }(\eta ^{\prime }y)\right\} _{y=\sqrt{u^{2}-k^{2}}},
\label{WFSingle2}
\end{eqnarray}%
where $I_{\nu }(z)$ is the modified Bessel function of the first kind. In
deriving this formula we have used the relations $K_{\nu }(ze^{\pm \pi
i})=e^{\mp \nu \pi i}K_{\nu }(z)\mp \pi iI_{\nu }(z)$. In the case $\beta >0$
the first and second terms on the right of (\ref{relcos}) have simple poles
at the points $i/\beta $ and $-i/\beta $ respectively. After the rotation of
the integration contour these points should be avoided by small semicircles
in the right half-plane. The integrals along these semicircles give
additional contributions which should be added to the expression (\ref%
{WFSingle2}) where now the integration over $u$ is understood in the sense
of the principal value. In order to not complicate formulae, in the
discussion below we will consider the case $\beta \leqslant 0$. Note that in
(\ref{WFSingle2}) the integration over the angular part of the vector $%
\mathbf{k}$ can be done explicitly and the expression for the Wightman
function takes the form%
\begin{eqnarray}
W(x,x^{\prime }) &=&W_{\mathrm{dS}}(x,x^{\prime })+\frac{(\eta \eta ^{\prime
})^{D/2}}{2(2\pi )^{(D+1)/2}\alpha ^{D-1}}\int_{0}^{\infty }dy\,y  \notag \\
&&\times \left\{ K_{\nu }(\eta y)\left[ I_{-\nu }(\eta ^{\prime }y)+I_{\nu
}(\eta ^{\prime }y)\right] +\left[ I_{\nu }(\eta y)+I_{-\nu }(\eta y)\right]
K_{\nu }(\eta ^{\prime }y)\right\}   \notag \\
&&\times \int_{0}^{\infty }dk\,k^{D-2}\frac{J_{(D-3)/2}(k|\mathbf{z}-\mathbf{%
z}^{\prime }|)}{(k|\mathbf{z}-\mathbf{z}^{\prime }|)^{(D-3)/2}}\frac{%
e^{-u(z^{D}+z^{D\prime })}}{u}\left. \frac{\beta u+1}{\beta u-1}\right\vert
_{u=\sqrt{y^{2}+k^{2}}},  \label{WFSingle4}
\end{eqnarray}%
where $J_{\nu }(z)$ is the Bessel function of the first kind.

For Dirichlet and Neumann boundary condition one has $\beta =0$\ and $%
1/\beta =0$, respectively, and the integral over $k$ in (\ref{WFSingle4}) is
evaluated by making use of the formula \cite{Prud86}
\begin{equation}
\int_{0}^{\infty }dk\,k^{D-2}\frac{J_{(D-3)/2}(k|\mathbf{z}-\mathbf{z}%
^{\prime }|)}{(k|\mathbf{z}-\mathbf{z}^{\prime }|)^{(D-3)/2}}\frac{e^{-\sqrt{%
y^{2}+k^{2}}(z^{D}+z^{D\prime })}}{\sqrt{y^{2}+k^{2}}}=\sqrt{\frac{2}{\pi }}%
y^{D/2-1}\frac{K_{D/2-1}(y\lambda )}{\lambda ^{D/2-1}},  \label{IntForm2}
\end{equation}%
with the notation $\lambda =\sqrt{|\mathbf{z}-\mathbf{z}^{\prime
}|^{2}+(z^{D}+z^{D\prime })^{2}}$. For the corresponding Wightman functions
we find%
\begin{eqnarray}
W^{\mathrm{(J)}}(x,x^{\prime }) &=&W_{\mathrm{dS}}(x,x^{\prime })-\delta ^{%
\mathrm{(J)}}\frac{(\eta \eta ^{\prime })^{D/2}\alpha ^{1-D}}{2(2\pi
)^{D/2+1}\lambda ^{D/2-1}}\int_{0}^{\infty }dy\,y^{D/2}K_{D/2-1}(y\lambda )
\notag \\
&&\times \left\{ K_{\nu }(\eta y)\left[ I_{-\nu }(\eta ^{\prime }y)+I_{\nu
}(\eta ^{\prime }y)\right] +\left[ I_{\nu }(\eta y)+I_{-\nu }(\eta y)\right]
K_{\nu }(\eta ^{\prime }y)\right\} ,  \label{WFSingle6}
\end{eqnarray}%
where J=D and J=N for Dirichlet and Neumann boundary conditions and $\delta
^{\mathrm{(D)}}=1$, $\delta ^{\mathrm{(N)}}=-1$.

\section{VEV of the field squared}

\label{sec:Phi2}

The VEV of the field squared is obtained from the Wightman function taking
the coincidence limit of the arguments. This limit is divergent and some
renormalization procedure is necessary. The important point here is that for
points away from the plate the divergences are contained in the part
corresponding to the boundary-free dS spacetime and the boundary induced
part is finite. As we have already extracted the first part, the
renormalization procedure is reduced to the renormalization of the
boundary-free dS part which is already done in the literature (see, for
instance, \cite{Bunc78,Cand75,Dowk76}). As a result the renormalized VEV of
the field squared is presented in the form%
\begin{equation}
\langle \varphi ^{2}\rangle =\langle \varphi ^{2}\rangle _{\mathrm{dS}%
}+\langle \varphi ^{2}\rangle _{\mathrm{pl}},  \label{phi2}
\end{equation}%
where $\langle \varphi ^{2}\rangle _{\mathrm{dS}}$ is the VEV\ in dS
spacetime without boundaries and the part
\begin{equation}
\langle \varphi ^{2}\rangle _{\mathrm{pl}}=\frac{4(4\pi )^{-(D+1)/2}}{\Gamma
((D-1)/2)\alpha ^{D-1}}\int_{0}^{\infty }dyy\left[ I_{\nu }(y)+I_{-\nu }(y)%
\right] K_{\nu }(y)\int_{y}^{\infty }dx\,G(x,y),  \label{phi21pl}
\end{equation}%
is induced by the plate. Here and in the discussion below we use the
notation
\begin{equation}
G(x,y)=(x^{2}-y^{2})^{(D-3)/2}\,e^{-2z^{D}x/\eta }\frac{\beta x/\eta +1}{%
\beta x/\eta -1}.  \label{Gxy}
\end{equation}%
Due to the maximal symmetry of the dS spacetime the VEV $\langle \varphi
^{2}\rangle _{\mathrm{dS}}$\ does not depend on the spacetime point. Note
that the boundary induced VEV is a function on the combinations $z^{D}/\eta $
and $\beta /\eta $ only. The ratio $z^{D}/\eta $ is the proper distance from
the plate measured in units of the curvature scale $\alpha $.

For a conformally coupled massless scalar field we have $\xi =\xi _{D}$, $%
\nu =1/2$ and the modified Bessel functions in (\ref{phi21pl}) are expressed
in terms of the elementary functions with the result $\left[ I_{\nu
}(y)+I_{-\nu }(y)\right] K_{\nu }(y)=1/y$. By making use of the formula
\begin{equation}
\int_{0}^{\infty }dy\int_{y}^{\infty }dx\,(x^{2}-y^{2})^{(D-3)/2}f(x)=\frac{%
\sqrt{\pi }\Gamma ((D-1)/2)}{2\Gamma (D/2)}\int_{0}^{\infty }dr\,r^{D-2}f(r),
\label{intrel}
\end{equation}%
the boundary induced part in the VEV\ of the field squared is presented in
the form%
\begin{equation}
\langle \varphi ^{2}\rangle _{\mathrm{pl}}=\frac{(\eta /\alpha )^{D-1}}{%
(4\pi )^{D/2}\Gamma (D/2)}\int_{0}^{\infty }dx\,x^{D-2}e^{-2xz^{D}}\frac{%
\beta x+1}{\beta x-1}.  \label{phi2Conf}
\end{equation}%
Formula (\ref{intrel}) is obtained changing the integration variable to $u=%
\sqrt{x^{2}-y^{2}}$ and introducing polar coordinates in the plane $(y,u)$.
Note that the result (\ref{phi2Conf}) could also be directly obtained by
using the fact that in the special case under consideration the problem is
conformally related to the corresponding problem for a Robin plate in the
Minkowski spacetime \cite{Rome02,Saha08Rev}. From this relation it follows
that $\langle \varphi ^{2}\rangle _{\mathrm{pl}}=[a(\eta )]^{1-D}\langle
\varphi ^{2}\rangle _{\mathrm{pl}}^{\mathrm{(M)}}$, with the scale factor $%
a(\eta )=\alpha /\eta $, which leads to the result (\ref{phi2Conf}).

Now let us investigate the behavior of the boundary induced VEV $\langle
\varphi ^{2}\rangle _{\mathrm{pl}}$ in the asymptotic regions of the ratio $%
z^{D}/\eta $. For small values of this ratio, $z^{D}/\eta \ll 1$, which
correspond to small proper distances from the plate with respect to the
curvature scale $\alpha $, we introduce new integration variables $%
z=xz^{D}/\eta $ and $u=yz^{D}/\eta $. The argument of the modified Bessel
functions becomes $u\eta /z^{D}$ and is large. By using the asymptotic
relation $\left[ I_{\nu }(y)+I_{-\nu }(y)\right] K_{\nu }(y)\approx 1/y$ for
large values $y$ and fixed $\nu $, we find%
\begin{equation}
\langle \varphi ^{2}\rangle _{\mathrm{pl}}\approx \frac{(\alpha z^{D}/\eta
)^{1-D}}{(4\pi )^{D/2}\Gamma (D/2)}\,\int_{0}^{\infty }dxx^{D-2}e^{-2x}\frac{%
\beta x/z^{D}+1}{\beta x/z^{D}-1},\;z^{D}/\eta \ll 1.  \label{phi2small}
\end{equation}%
Hence, in the limit under consideration to the leading order the VEV
coincides with the corresponding result for a conformally coupled massless
scalar. For fixed value $z^{D}$ this limit corresponds to early stages of
the cosmological expansion. As the boundary-free part $\langle \varphi
^{2}\rangle _{\mathrm{dS}}$ is a constant, we conclude that at small proper
distances from the plate the total VEV of the field squared is dominated by
the boundary induced part. From (\ref{phi2small}) it follows that for
Dirichlet boundary condition ($\beta =0$) this part is negative. For
non-Dirichlet boundary conditions, assuming that $z^{D}\ll |\beta |$, near
the plate the VEV of the field squared is positive.

In the opposite limit of large values for the ratio $z^{D}/\eta $ (large
proper distances from the plate with respect to the curvature scale $\alpha $%
), the main contribution in the integral over $x$ in (\ref{phi21pl}) comes
from the region near the lower limit of the integration and for positive
values of the parameter $\nu $ to the leading order one has%
\begin{equation}
\langle \varphi ^{2}\rangle _{\mathrm{pl}}\approx \frac{\delta _{\beta
}\Gamma ((D+1)/2-2\nu )}{2^{1-2\nu }(2\pi )^{(D+1)/2}\alpha ^{D-1}}\frac{%
\Gamma (\nu )}{\Gamma (1-\nu )}\left( \frac{\eta }{2z^{D}}\right) ^{D-2\nu
},\;z^{D}/\eta \gg 1,  \label{phi2largereal}
\end{equation}%
where $\delta _{\beta }=1$ for $1/\beta =0$ and $\delta _{\beta }=-1$ for $%
1/\beta \neq 0$. In the case $\nu =0$ the VEV\ of the field squared behaves
as $\langle \varphi ^{2}\rangle _{\mathrm{pl}}\propto (\eta /z^{D})^{D}\ln
(z^{D}/\eta )$. For fixed values of $z^{D}$ this limit corresponds to late
stages of the cosmological expansion. For imaginary values of $\nu $, by
using the relation
\begin{equation}
K_{\nu }(y)\left[ I_{-\nu }(y)+I_{\nu }(y)\right] \approx \frac{\pi }{\sinh
(|\nu |\pi )}{\mathrm{Im}}\left[ \frac{(y/2)^{-2i|\nu |}}{\Gamma
^{2}(1-i|\nu |)}\right] ,  \label{IKsmall}
\end{equation}%
for small values of $y$, in the leading order we find
\begin{equation}
\langle \varphi ^{2}\rangle _{\mathrm{pl}}\approx \frac{\delta _{\beta
}(2\pi )^{-(D+1)/2}}{\alpha ^{D-1}(2z^{D}/\eta )^{D}}\frac{\pi B_{0}}{\sinh
(|\nu |\pi )}\sin [2|\nu |\ln (2z^{D}/\eta )+\phi _{0}].
\label{phi2largeimag}
\end{equation}%
In this formula the constants $B_{0}$ and $\phi _{0}$ are defined by the
relation
\begin{equation*}
B_{0}e^{i\phi _{0}}=\frac{2^{2i|\nu |}\Gamma ((D+1)/2-2i|\nu |)}{\Gamma
^{2}(1-i|\nu |)}.
\end{equation*}%
As we see, in this case the behavior of the boundary induced VEV as a
function of the proper distance from the plate is oscillatory damping. Note
that in terms of the synchronous time coordinate the expression in the phase
of (\ref{phi2largeimag}) is written in the form $\ln (2z^{D}/\eta )=t/\alpha
+\ln (2z^{D}/\alpha )$. From the asymptotic expressions given above we
conclude that at large proper distances from the plate the VEV\ of the field
squared is dominated by the boundary-free part $\langle \varphi ^{2}\rangle
_{\mathrm{dS}}$.

\section{Energy-momentum tensor}

\label{sec:EMT}

Now we turn to the investigation of the VEV for the energy-momentum tensor.
Having the Wightman function we can evaluate this VEV by making use of the
formula%
\begin{equation}
\langle T_{ik}\rangle =\lim_{x^{\prime }\rightarrow x}\partial _{i}\partial
_{k}^{\prime }W(x,x^{\prime })+\left[ \left( \xi -1/4\right) g_{ik}\nabla
_{l}\nabla ^{l}-\xi \nabla _{i}\nabla _{k}-\xi R_{ik}\right] \langle \varphi
^{2}\rangle ,  \label{emtvev1}
\end{equation}%
where $R_{ik}=Dg_{ik}/\alpha ^{2}$ is the Ricci tensor for the dS spacetime.
Similar to the case of the field squared, the VEV of the energy-momentum
tensor is presented in the decomposed form
\begin{equation}
\langle T_{ik}\rangle =\langle T_{ik}\rangle _{\mathrm{dS}}+\langle
T_{ik}\rangle _{\mathrm{pl}},  \label{TikDecomp}
\end{equation}%
where $\langle T_{ik}\rangle _{\mathrm{dS}}$ is the corresponding VEV\ in dS
spacetime without boundaries and $\langle T_{ik}\rangle _{\mathrm{pl}}$ is
induced by the plate. For points away the plate the latter is finite and the
renormalization is necessary for the boundary-free part only. Due to the dS
invariance of the Bunch-Davies vacuum this part is proportional to the
metric tensor with a constant coefficient and is well investigated in the
literature \cite{Bunc78,Cand75,Dowk76}. For this reason in the discussion
below we will be focused on the boundary induced part.

In order to evaluate the boundary induced part in the VEV of the
energy-momentum tensor we need the expression for the covariant
d'Alembertian acted on the VEV of the field squared:%
\begin{eqnarray}
\nabla _{l}\nabla ^{l}\langle \varphi ^{2}\rangle _{\mathrm{pl}} &=&\frac{%
4(4\pi )^{-(D+1)/2}}{\Gamma ((D-1)/2)\alpha ^{D+1}}\int_{0}^{\infty
}dy\,y^{3-D}\int_{y}^{\infty }dx\,G(x,y)  \notag \\
&&\times \left( \partial _{y}^{2}+\frac{1-D}{y}\partial _{y}-\frac{4x^{2}}{%
y^{2}}\right) \tilde{I}_{\nu }(y)\tilde{K}_{\nu }(y),  \label{Dalamb}
\end{eqnarray}%
where the notations%
\begin{equation}
\tilde{K}_{\nu }(z)=z^{D/2}K_{\nu }(z),\;\tilde{I}_{\nu }(z)=z^{D/2}\left[
I_{\nu }(z)+I_{-\nu }(z)\right]   \label{Ktilde}
\end{equation}%
are introduced. Now, by using the formula (\ref{emtvev1}), after long
calculations, we present the VEV\ of the energy-momentum tensor in the form
\begin{equation}
\langle T_{k}^{i}\rangle _{\mathrm{pl}}=\frac{4(4\pi )^{-(D+1)/2}}{\Gamma
((D-1)/2)\alpha ^{D+1}}\int_{0}^{\infty }dy\,y^{1-D}\int_{y}^{\infty
}dx\,G(x,y)\left[ F_{k}^{i}(y)+x^{2}F_{k}^{i}\tilde{I}_{\nu }(y)\tilde{K}%
_{\nu }(y)\right] ,  \label{Tii1}
\end{equation}%
where the function $G(x,y)$ is defined by (\ref{Gxy}). In formula (\ref{Tii1}%
) we have introduced the notations (no summation over $i$)
\begin{eqnarray}
F_{0}^{0}(y) &=&y^{2}\tilde{I}_{\nu }^{\prime }(y)\tilde{K}_{\nu }^{\prime
}(y)-\left[ \frac{1}{4}y^{2}\partial _{y}^{2}+D(\xi -\xi _{D})y\partial
_{y}+\xi D\right] \tilde{I}_{\nu }(y)\tilde{K}_{\nu }(y),  \notag \\
F_{i}^{i}(y) &=&F_{D}^{D}(y)+\frac{y^{2}}{D-1}\tilde{I}_{\nu }(y)\tilde{K}%
_{\nu }(y),\;i=1,2,\ldots ,D-1,  \label{Fi} \\
F_{D}^{D}(y) &=&\left\{ \left( \xi -\frac{1}{4}\right) y^{2}\partial
_{y}^{2}+\left[ \xi (2-D)+\frac{D-1}{4}\right] y\partial _{y}-\xi D\right\}
\tilde{I}_{\nu }(y)\tilde{K}_{\nu }(y),  \notag \\
F_{0}^{D}(y) &=&2x\left[ \left( \xi -1/4\right) y\partial _{y}+\xi \right]
\tilde{I}_{\nu }(y)\tilde{K}_{\nu }(y),  \notag
\end{eqnarray}%
and%
\begin{equation}
F_{0}^{0}=1-4\xi ,\;F_{i}^{i}=1-4\xi -\frac{1}{D-1},\;F_{0}^{D}=F_{D}^{D}=0.
\label{F0n}
\end{equation}%
As in the case of the field squared, the VEVs of the energy-momentum tensor
are functions of the ratios $z^{D}/\eta $ and $\beta /\eta $ only. Note
that, by using the well-known properties of the modified Bessel functions,
the function $F_{0}^{0}(y)$ can also be written in the form%
\begin{equation}
F_{0}^{0}(y)=\left[ (y^{2}/4)\partial _{y}^{2}-D(\xi +\xi _{D})y\partial
_{y}-y^{2}+D^{2}\xi +m^{2}\alpha ^{2}\right] \tilde{I}_{\nu }(y)\tilde{K}%
_{\nu }(y).  \label{F0}
\end{equation}%
As we see, the vacuum energy-momentum tensor is non-diagonal with the
off-diagonal component $\langle T_{0}^{D}\rangle _{\mathrm{pl}}$ which
describes the energy flux along the direction normal to the plate. This flux
can be either positive or negative (see examples plotted in figures below
for Dirichlet boundary condition). Note that this type of the energy flux
appears also in the geometry of a cosmic string on background of
Friedmann-Robertson-Walker and dS spacetimes \cite{Davi88,Beze09}.

It can be checked that the boundary induced parts in the VEV\ of the
energy-momentum tensor satisfy the trace relation%
\begin{equation}
\langle T_{i}^{i}\rangle _{\mathrm{pl}}=\left[ D(\xi -\xi _{D})\nabla
_{i}\nabla ^{i}+m^{2}\right] \langle \varphi ^{2}\rangle _{\mathrm{pl}}.
\label{tracerel}
\end{equation}%
In particular, the plate induced part is traceless for a conformally coupled
massless scalar field. Trace anomaly is contained in the boundary-free part
only.

For a conformally coupled massless field we have $\nu =1/2$. In this case $%
\tilde{I}_{\nu }(y)\tilde{K}_{\nu }(y)=y^{D-1}$, and it can be checked that $%
F_{0}^{D}(y)=F_{D}^{D}(y)=0$. For the energy density the corresponding
expression takes the form
\begin{eqnarray}
\langle T_{0}^{0}\rangle _{\mathrm{pl}} &=&\frac{4(4\pi )^{-(D+1)/2}\eta
^{D+1}}{D\Gamma ((D-1)/2)\alpha ^{D+1}}\int_{0}^{\infty
}dy\,\int_{y}^{\infty }dx\,(x^{2}-y^{2})^{(D-3)/2}\,  \notag \\
&&\times e^{-2z^{D}x}\frac{\beta x+1}{\beta x-1}(x^{2}-Dy^{2}),
\label{T00Conf}
\end{eqnarray}%
and for the stresses along the directions parallel to the plate we have (no
summation over $i$) $\langle T_{i}^{i}\rangle _{\mathrm{pl}}=-(D-1)\langle
T_{0}^{0}\rangle _{\mathrm{pl}}$, $i=1,2,\ldots ,D-1$. Introducing in (\ref%
{T00Conf}) a new integration variable $z=\sqrt{x^{2}-y^{2}}$ and polar
coordinates $y=r\cos \theta $, $z=r\sin \theta $, we can see that the
integral over $\theta $ vanishes. Hence, for a\ conformally coupled field
the plate induced part vanishes. This result can also be obtained by using
the conformal relation between the problem under consideration and the
corresponding problem in the Minkowski spacetime. In the latter geometry the
Casimir energy-momentum tensor for a conformally coupled massless scalar
field vanishes.

The general formulae for the VEV of the energy-momentum tensor are
simplified in the asymptotic regions of the ratio $z^{D}/\eta $. For small
values of this parameter, corresponding to small proper distances from the
plate, by using the asymptotic formulae for the modified Bessel functions
for large values of the argument, we find the following asymptotic behavior:%
\begin{eqnarray}
\langle T_{l}^{l}\rangle _{\mathrm{pl}} &\approx &-\frac{4(\xi -\xi _{D})}{%
(\alpha z^{D}/\eta )^{D+1}}\mathcal{I}_{D}(\beta /z^{D}),\;l=0,1,\ldots ,D-1,
\notag \\
\langle T_{0}^{D}\rangle _{\mathrm{pl}} &\approx &\frac{2D(\xi -\xi _{D})}{%
\alpha (\alpha z^{D}/\eta )^{D}}\mathcal{I}_{D-1}(\beta /z^{D}),\;\langle
T_{D}^{D}\rangle _{\mathrm{pl}}\approx -\frac{D(\xi -\xi _{D})}{\alpha
^{2}(\alpha z^{D}/\eta )^{D-1}}\mathcal{I}_{D-2}(\beta /z^{D}),
\label{emtAsLarge}
\end{eqnarray}%
where we have introduced the notation%
\begin{equation*}
\mathcal{I}_{\mu }(z)=\frac{(4\pi )^{-D/2}}{\Gamma \left( D/2\right) }%
\int_{0}^{\infty }dx\,x^{\mu }e^{-2x}\frac{zx+1}{zx-1}.
\end{equation*}%
Hence, the VEVs diverge on the boundary. The ratio $\alpha z^{D}/\eta $ in
these formulae is the proper distance from the plate and the leading
divergences in the components $\langle T_{l}^{l}\rangle _{\mathrm{pl}}$, $%
l=0,1,\ldots ,D-1$, are the same as those in the corresponding problem on
the Minkowski bulk. For the latter problem the components $\langle
T_{0}^{D}\rangle _{\mathrm{pl}}$ and $\langle T_{D}^{D}\rangle _{\mathrm{pl}%
} $ vanish. In the case of the dS bulk near the plate we have $|\langle
T_{D}^{D}\rangle _{\mathrm{pl}}|\ll |\langle T_{0}^{D}\rangle _{\mathrm{pl}%
}|\ll |\langle T_{0}^{0}\rangle _{\mathrm{pl}}|$. Note that for a
conformally coupled field the leading terms vanish and the next terms in the
corresponding asymptotic expansions should be taken into account. The limit
under consideration corresponds to points near the plate or to late stages
of the cosmological expansion. In this limit the total VEV of the
energy-momentum tensor is dominated by the boundary induced part. Note that
the function $\mathcal{I}_{\mu }(\beta /z^{D})$ is negative for Dirichlet
boundary condition ($\beta =0$) and is positive for non-Dirichlet boundary
conditions assuming that $z^{D}\ll |\beta |$.

Now let us consider the behavior of the vacuum energy-momentum tensor in the
asymptotic region $z^{D}/\eta \gg 1$. In this limit the main contribution in
the integral over $x$ in (\ref{Tii1}) comes from the region of the
integration near the lower limit. As for the field squared, we consider the
cases of the real and imaginary values $\nu $ separately. For real $\nu >0$,
by using the asymptotic formulae for the modified Bessel functions for small
values of the argument, in the leading order we find (no summation over $i$)%
\begin{equation}
\langle T_{i}^{i}\rangle _{\mathrm{pl}}\approx -\frac{\Gamma (1+\nu )}{%
\Gamma (1-\nu )}\frac{2^{2\nu }\delta _{\beta }{}\Gamma ((D+1)/2-2\nu
)F_{\nu }^{(i)}}{(2\pi )^{(D+1)/2}\alpha ^{D+1}(2z^{D}/\eta )^{D-2\nu }},
\label{Tii1largezD}
\end{equation}%
for the diagonal components and
\begin{equation}
\langle T_{0}^{D}\rangle _{\mathrm{pl}}\approx \frac{2^{2\nu }\Gamma (\nu )}{%
\Gamma (1-\nu )}\frac{\delta _{\beta }{}\Gamma ((D+3)/2-2\nu )F_{\nu }^{(D)}%
}{(2\pi )^{(D+1)/2}\alpha ^{D+1}(2z^{D}/\eta )^{D+1-2\nu }},
\label{T0D1largezD}
\end{equation}%
for the off-diagonal component with $\delta _{\beta }$ defined after formula
(\ref{phi2largereal}). In these expressions the notations%
\begin{eqnarray}
F_{\nu }^{(0)} &=&-D\xi +\frac{D-2\nu }{4}-\frac{m^{2}\alpha ^{2}}{2\nu },\;
\notag \\
F_{\nu }^{(l)} &=&\xi (D-2\nu +1)-\frac{D-2\nu }{4},\;l=1,\ldots ,D,
\label{Fnul}
\end{eqnarray}%
are introduced. As we see, in the limit under consideration the vacuum
stresses are isotropic and $|\langle T_{0}^{D}\rangle _{\mathrm{pl}}|\ll
|\langle T_{0}^{0}\rangle _{\mathrm{pl}}|$. For $\nu =0$ one has the
asymptotic expressions%
\begin{equation}
\langle T_{0}^{0}\rangle _{\mathrm{pl}}\approx -\frac{2Dz^{D}}{\eta (D+1)}%
\langle T_{0}^{D}\rangle _{\mathrm{pl}}\approx m^{2}\alpha ^{2}\frac{2\delta
_{\beta }\Gamma ((D+1)/2)\ln (2z^{D}/\eta )}{(2\pi )^{(D+1)/2}\alpha
^{D+1}(2z^{D}/\eta )^{D}},  \label{T00nu0}
\end{equation}%
and for the diagonal stresses we have (no summation over $i$) $|\langle
T_{i}^{i}\rangle _{\mathrm{pl}}|\ll |\langle T_{0}^{0}\rangle _{\mathrm{pl}%
}| $, $i=1,\ldots ,D$.

For imaginary values $\nu $ we use the asymptotic relation (\ref{IKsmall}).
To the leading order this gives the results (no summation over $l$)%
\begin{eqnarray}
\langle T_{l}^{l}\rangle _{\mathrm{pl}} &\approx &-\frac{\delta _{\beta
}{}(2\pi )^{-(D+1)/2}}{(2z^{D}/\eta )^{D}\alpha ^{D+1}}\frac{2|\nu |\pi
B^{(l)}}{\sinh (|\nu |\pi )}\sin \left[ 2|\nu |\ln (2z^{D}/\eta )+\phi _{l}%
\right] ,  \notag \\
\langle T_{0}^{D}\rangle _{\mathrm{pl}} &\approx &\frac{2\delta _{\beta
}{}(2\pi )^{-(D+1)/2}}{(2z^{D}/\eta )^{D+1}\alpha ^{D+1}}\frac{\pi B^{(0D)}}{%
\sinh (|\nu |\pi )}\sin \left[ 2|\nu |\ln (2z^{D}/\eta )+\phi _{0D}\right] ,
\label{T0Dimagnu}
\end{eqnarray}%
where the constants are defined by the relations%
\begin{eqnarray}
B^{(l)}e^{i\phi _{l}} &=&2^{2i|\nu |}\frac{i\Gamma ((D+1)/2-2i|\nu |)}{%
\Gamma ^{2}(1-i|\nu |)}F_{i|\nu |}^{(l)},  \notag \\
B^{(0D)}e^{i\phi _{0D}} &=&2^{2i|\nu |}\frac{\Gamma ((D+3)/2-2i|\nu |)}{%
\Gamma ^{2}(1-i|\nu |)}F_{i|\nu |}^{(D)}.  \label{B0D}
\end{eqnarray}%
In this case the damping of the Casimir densities is oscillatory. For a
fixed value of $z^{D}$ the oscillating period in terms of the synchronous
time coordinate is equal to $\alpha /(2\pi )$.

\section{Dirichlet and Neumann boundary conditions}

\label{sec:Dir}

The formulae given above for the boundary induced parts in the VEVs of the
field squared and the energy-momentum tensor are further simplified in the
special cases of Dirichlet and Neumann boundary conditions. By using the
formula
\begin{equation}
\int_{y}^{\infty }dx\,(x^{2}-y^{2})^{(D-3)/2}\,e^{-2z^{D}x/\eta }=2^{D/2-1}%
\frac{\Gamma ((D-1)/2)}{y^{2-D}\sqrt{\pi }}\frac{K_{D/2-1}(2z^{D}y/\eta )}{%
(2z^{D}y/\eta )^{D/2-1}},  \label{IntForm4}
\end{equation}%
for the VEV\ of the field squared from (\ref{phi21pl}) one finds%
\begin{equation}
\langle \varphi ^{2}\rangle _{\mathrm{pl}}^{\mathrm{(J)}}=-\frac{2\delta ^{%
\mathrm{(J)}}\alpha ^{1-D}}{(2\pi )^{D/2+1}}\int_{0}^{\infty }dyy^{D/2}\left[
I_{\nu }(y)+I_{-\nu }(y)\right] K_{\nu }(y)\frac{K_{D/2-1}(2z^{D}y/\eta )}{%
(2z^{D}y/\eta )^{D/2-1}},  \label{phi2Dir}
\end{equation}%
with J=D,N for Dirichlet and Neumann boundary conditions and the definition
for $\delta ^{\mathrm{(J)}}$ is given after formula (\ref{WFSingle6}).

In the similar way, for the off-diagonal component of the VEV of the
energy-momentum tensor we have%
\begin{equation}
\langle T_{0}^{D}\rangle _{\mathrm{pl}}^{\mathrm{(J)}}=\frac{4\delta ^{%
\mathrm{(J)}}\alpha ^{-D-1}}{(2\pi )^{D/2+1}}\int_{0}^{\infty }dy\,\frac{%
K_{D/2}(2z^{D}y/\eta )}{(2z^{D}y/\eta )^{D/2-1}}\left[ \left( 1/4-\xi
\right) y\partial _{y}-\xi \right] \tilde{I}_{\nu }(y)\tilde{K}_{\nu }(y).
\label{T0DDir}
\end{equation}%
For the diagonal components we find the expressions (no summation over $i$)%
\begin{eqnarray}
\langle T_{i}^{i}\rangle _{\mathrm{pl}}^{\mathrm{(J)}} &=&-\frac{2\delta ^{%
\mathrm{(J)}}\alpha ^{-D-1}}{(2\pi )^{D/2+1}}\int_{0}^{\infty }dy\,y\left\{
\frac{F_{i}^{i}(y)}{y^{2}}\frac{K_{D/2-1}(z)}{z^{D/2-1}}\right.   \notag \\
&&\left. +F_{i}^{i}\tilde{I}_{\nu }(y)\tilde{K}_{\nu }(y)\left[ \frac{%
K_{D/2+1}(z)}{z^{D/2-1}}-\frac{K_{D/2}(z)}{z^{D/2}}\right] \right\}
_{z=2z^{D}y/\eta },  \label{TiiDir}
\end{eqnarray}%
with the notations (\ref{F0n}). The corresponding asymptotic expressions are
directly obtained from those given above for the general Robin case.

\begin{figure}[tbph]
\begin{center}
\begin{tabular}{cc}
\epsfig{figure=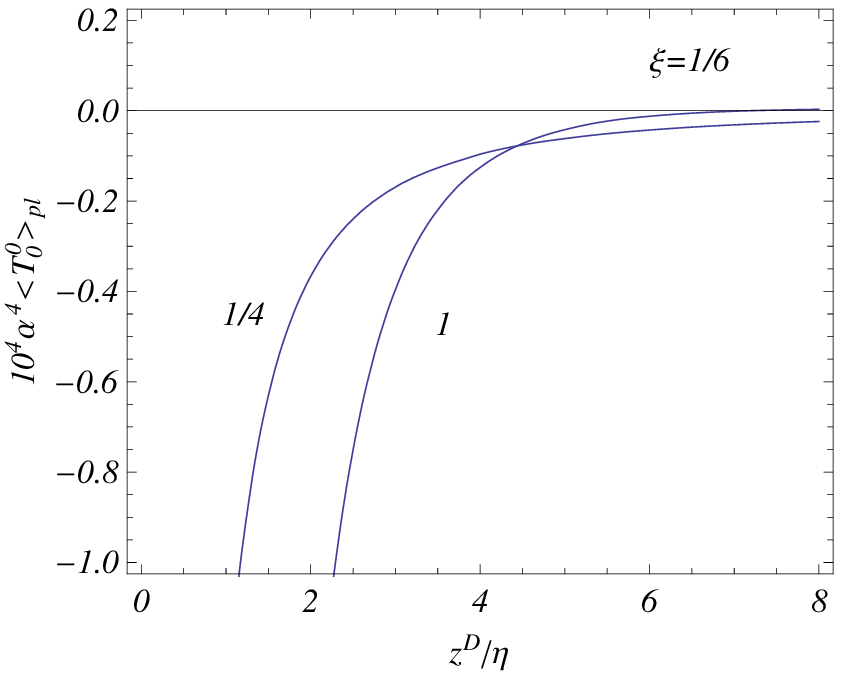,width=7.cm,height=6.cm} & \quad %
\epsfig{figure=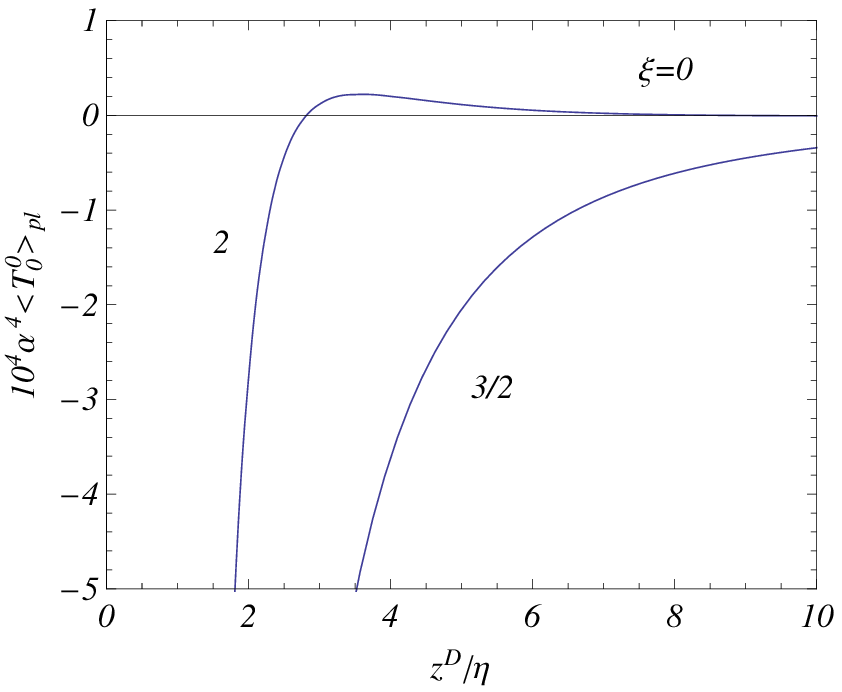,width=7.cm,height=6cm}%
\end{tabular}%
\end{center}
\caption{The boundary induced part in the VEV of the energy density for $D=3$
scalar field with Dirichlet boundary condition as a function of the ratio $%
z^D/\protect\eta $. The left/right panel corresponds to fields with
conformal/minimal coupling. The numbers near the curves are the values of
the parameter $\protect\alpha m$. }
\label{fig1}
\end{figure}

\begin{figure}[tbph]
\begin{center}
\begin{tabular}{cc}
\epsfig{figure=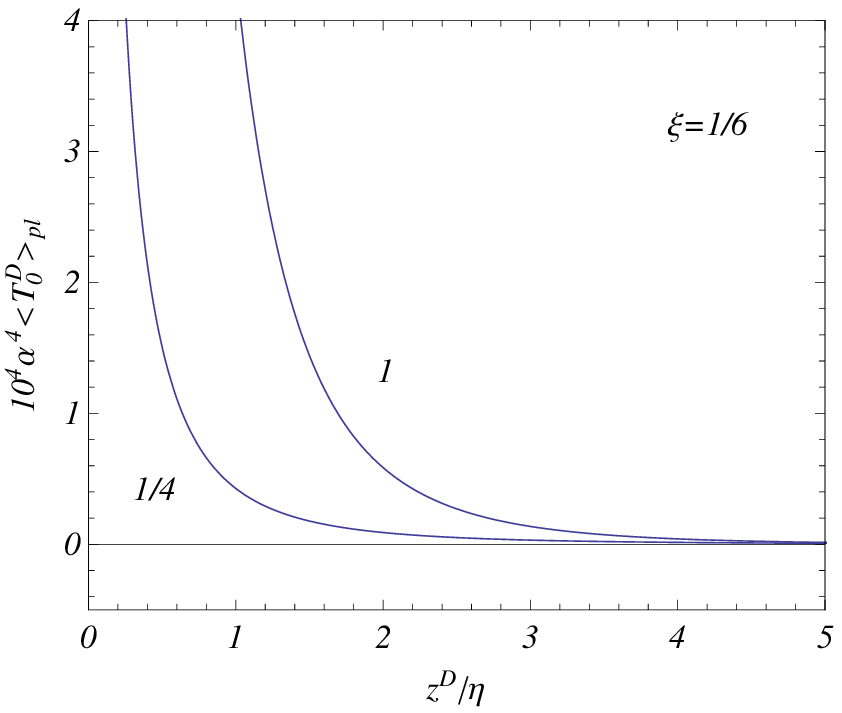,width=7.cm,height=6.cm} & \quad %
\epsfig{figure=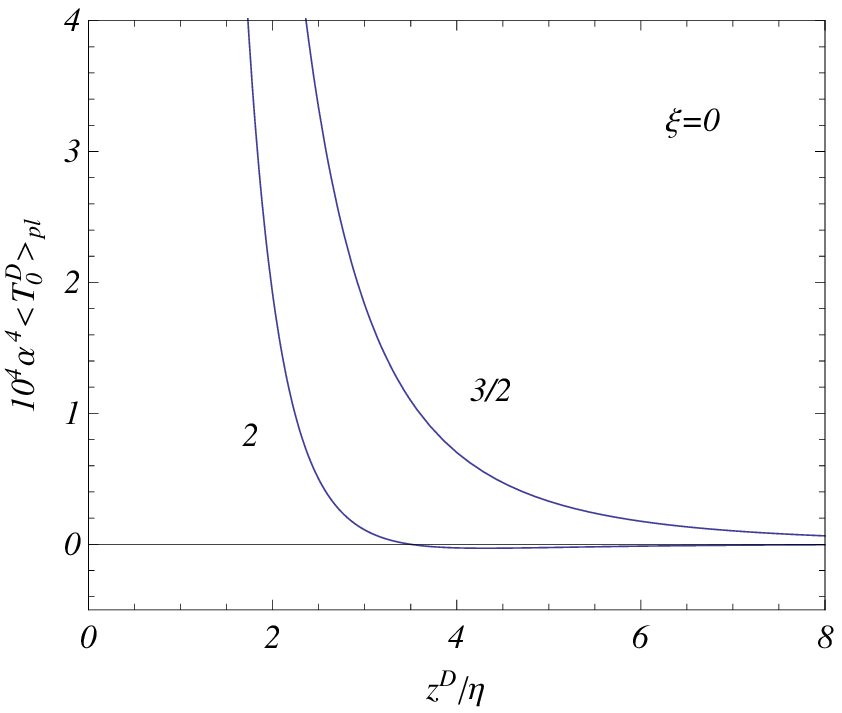,width=7.cm,height=6cm}%
\end{tabular}%
\end{center}
\caption{The same as in figure \protect\ref{fig1} for the energy flux.}
\label{fig2}
\end{figure}

In figures \ref{fig1} and \ref{fig2} for conformally and minimally coupled $%
D=3$ Dirichlet scalar fields we have plotted the plate induced parts in the
VEVs of the energy density and the energy flux along the direction normal to
the plate as functions of the ratio $z^{D}/\eta $. Recall that the latter is
the proper distance from the plate measured in units of the dS curvature
scale $\alpha $. The numbers near the curves correspond to the values of the
parameter $m\alpha $. We have taken these values in the way to have both
real and imaginary values for the parameter $\nu $. In the first case
(graphs for $\xi =1/6$, $m\alpha =1/4$ and $\xi =0$, $m\alpha =3/2$) the
VEVs decay at large distances monotonically, whereas in the second case
(graphs for $\xi =1/6$, $m\alpha =1$ and $\xi =0$, $m\alpha =2$) the
corresponding behavior is damping oscillatory (see formulae (\ref%
{Tii1largezD}), (\ref{T0D1largezD}), (\ref{T0Dimagnu}) for the
corresponding asymptotics). In the case of a conformally coupled
field with $m\alpha =1$ the first zero is at $z^{D}/\eta \approx
7.19$ for the energy density and at $z^{D}/\eta \approx 9.64$ for
the energy flux. For a minimally coupled field with $m\alpha =2$
the first two zeroes are located at $z^{D}/\eta \approx 2.82,$
$8.67$ for the energy density and at $z^{D}/\eta \approx 3.5,11.3$
for the energy flux. The energy flux near the plate is positive
for both conformally and minimally coupled scalars. For Neumann
boundary conditions the VEVs have opposite signs.

\begin{figure}[tbph]
\begin{center}
\epsfig{figure=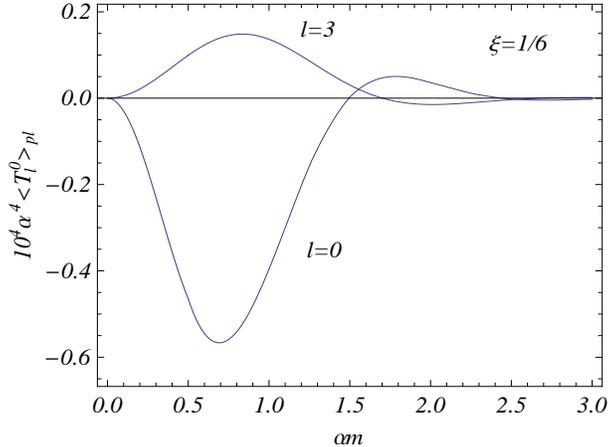,width=8cm,height=6.cm}
\end{center}
\caption{The boundary induced VEVs of the energy density and flux
versus $\alpha m $ in the case of $D=3$ conformally coupled scalar
field with Dirichlet boundary condition for the fixed distance
from the plate $z^D/\protect\eta =3$.} \label{fig3}
\end{figure}

It is also of interest to see the dependence of the boundary
induced parts in the VEVs on the mass of the field. In figure
\ref{fig3} we have presented this dependence for the energy
density and flux in the case of a conformally coupled field for
the fixed distance from the plate corresponding to $z^{D}/\eta
=3$. For large values of the mass the corresponding asymptotic
behavior is described by formulae (\ref{T0Dimagnu}).

\section{Conclusion}

\label{sec:Conc}

In the present paper we have investigated the VEVs of the field squared and
the energy-momentum tensor for a massive scalar field with general curvature
coupling parameter induced by a plane boundary in dS spacetime. The Robin
boundary condition was assumed on the boundary. As the first step we have
constructed the corresponding positive frequency Wightman function. This
function is presented in the form of the sum of boundary-free and boundary
induced parts. For points away from the boundary the latter is finite in the
coincidence limit of the arguments and can be directly used for the
evaluation of the VEVs of the field squared and the energy-momentum tensor.
These VEVs are decomposed into boundary-free dS and plate induced parts. Due
to the maximal symmetry of dS spacetime and the dS invariance of the
Bunch-Davies vacuum the boundary-free parts do not depend on the spacetime
point. These parts are well investigated in the literature and we were
focused on the plate induced parts. The latter are given by formula (\ref%
{phi21pl}) for the field squared and by formula (\ref{Tii1}) for the
energy-momentum tensor and they are functions on the combinations $%
z^{D}/\eta $ and $\beta /\eta $ only. The first of these ratios is the
proper distance from the plate measured in units of the curvature scale $%
\alpha $. An interesting feature is that the vacuum energy-momentum tensor
is non-diagonal with the off-diagonal component corresponding to the energy
flux along the direction normal to the plate. In dependence of the
parameters, this flux can be either positive or negative. For special case
of a conformally coupled massless scalar field the general formula for the
VEV of the field squared is simplified to (\ref{phi2Conf}) and the plate
induced energy-momentum tensor vanishes. These results can also be obtained
from the corresponding flat spacetime results by using the conformal
relation between the geometries.

For general values of the curvature coupling parameter the formulae for the
plate induced VEVs are further simplified in the asymptotic regions of small
and large proper distances from the plate. In the first case the leading
term in the asymptotic expansions have the form (\ref{phi2small}) for the
field squared and the form (\ref{emtAsLarge}) for the energy-momentum
tensor. Near the plate we have $|\langle T_{D}^{D}\rangle _{\mathrm{pl}}|\ll
|\langle T_{D}^{0}\rangle _{\mathrm{pl}}|\ll |\langle T_{0}^{0}\rangle _{%
\mathrm{pl}}|$ and the total VEV of the energy-momentum tensor is dominated
by the boundary induced part. At large proper distances from the plate the
behavior of the plate induced parts is qualitatively different for real and
pure imaginary values of the parameter $\nu $. In the first case these parts
decay monotonically as $(z^{D}/\eta )^{2\nu -D}$ for the field squared and
for the diagonal components of the energy-momentum tensor and like $%
(z^{D}/\eta )^{2\nu -D-1}$ for the off-diagonal component $\langle
T_{0}^{D}\rangle _{\mathrm{pl}}$. In this limit the diagonal vacuum stresses
are isotropic. For imaginary values of the parameter $\nu $ the asymptotic
behavior of the plate induced VEVs at large proper distances from the plate
is described by formulae (\ref{phi2largeimag}) and (\ref{T0Dimagnu}) and is
damping oscillatory. For a fixed value of $z^{D}$ the oscillating period in
terms of the synchronous time coordinate is equal to $\alpha /(2\pi )$.

The special cases of Dirichlet and Neumann boundary conditions are
considered in section \ref{sec:Dir}. For these cases the formulae for the
expectation values are further simplified to (\ref{phi2Dir})-(\ref{TiiDir})
and they have opposite signs for Dirichlet and Neumann scalars.

\section*{Acknowledgments}

The work was supported by the Armenian Ministry of Education and Science
Grant No. 119. A.A.S. gratefully acknowledges the hospitality of the INFN
Laboratori Nazionali di Frascati (Frascati, Italy) where part of this work
was done.

\end{document}